\documentclass[11pt]{article}
                                                                                
\usepackage{fullpage}
\usepackage{latexsym}
\usepackage{amsmath}
\usepackage{amssymb}
\usepackage{amsfonts}

\def\01{\{0,1\}}
\newcommand{\ceil}[1]{\lceil{#1}\rceil}

\newcommand{\eps}{\varepsilon}
\newcommand{\polylog}{\mbox{polylog}}
\newcommand{\ket}[1]{|#1\rangle}
\newcommand{\bra}[1]{\langle#1|}
\newcommand{\outp}[2]{|#1\rangle\langle#2|}
\newcommand{\ketbra}[2]{|#1\rangle\langle#2|}

\newcommand{\inp}[2]{\langle{#1}|{#2}\rangle} 

\newcommand{\norm}[1]{\mbox{$\parallel{#1}\parallel$}}

\newtheorem{definition}{Definition}
\newtheorem{theorem}{Theorem}
\newtheorem{lemma}{Lemma}

\newtheorem{corollary}{Corollary}

\newenvironment{proof}[1][Proof.]{
        \par
        \noindent \textbf{#1}
}{
        \nobreak\leavevmode
        \hfill $\Box$\par\bigskip
}

\newcommand{\cemph}[1]{#1}
 
\begin{document}
\title{Improved Lower Bounds for Locally Decodable Codes and Private Information Retrieval}
\author{Stephanie Wehner\thanks{Supported by EU project RESQ IST-2001-37559 and NWO Vici grant 2004-2009.}\\
        CWI, Amsterdam\\
        wehner@cwi.nl
        \and
        Ronald de~Wolf$^\star$\\
        CWI, Amsterdam\\
        rdewolf@cwi.nl
}
\maketitle

\begin{abstract}
We prove new lower bounds for \emph{locally decodable codes}
and \emph{private information retrieval}. We show that
a 2-query LDC encoding $n$-bit strings over an $\ell$-bit alphabet, 
where the decoder only uses $b$ bits of each queried position, 
needs code length
$
m=\exp\left(\Omega\left(\frac{n}{2^b\sum_{i=0}^b{\ell \choose i}}\right)\right).
$
Similarly, a 2-server PIR scheme with an $n$-bit database and
$t$-bit queries, where the user only needs $b$ bits from each of the two 
$\ell$-bit answers, unknown to the servers, satisfies
$
t=\Omega\left(\frac{n}{2^b\sum_{i=0}^b{\ell \choose i}}\right).
$
This implies that several known PIR schemes are close to optimal.
Our results generalize those of Goldreich et~al.~\cite{gkst:lowerpir},
who proved roughly the same bounds for \emph{linear} LDCs and PIRs.
Like earlier work by Kerenidis and de Wolf~\cite{kerenidis&wolf:qldcj},
our classical bounds are proved using quantum computational techniques.
In particular, we give a tight analysis of how well a 2-input function
can be computed from a quantum superposition of both inputs.\\[2mm]
\end{abstract}
\section{Introduction}
\subsection{Locally decodable codes}

Error correcting codes allow reliable transmission and storage of
information in noisy environments. Such codes often have the 
disadvantage that one has to read almost the entire codeword, even 
if one is only interested in a small part of the encoded information.
A \emph{locally decodable} code $\cemph{C}:\01^n\rightarrow\Sigma^m$ 
over alphabet $\Sigma$ is an error-correcting code that allows 
efficient decoding of individual bits of the encoded information:
given any string $y$ that is sufficiently close to the real codeword 
$\cemph{C}(x)$, we can probabilistically recover any bit $x_i$ of 
the original input $x$, while only looking at $k$ positions of $y$. 
The code length $m$ measures the cost of the encoding, 
while $k$ measures the efficiency of decoding individual bits.
Such codes have had a number of applications in recent computer science
research, including PCPs and worst-case to average-case reductions.
One can also think of applications encoding a large chunk 
of data in order to protect it from noise, where we are only interested
in extracting small pieces at a time. Imagine for example an encoding
of all books in a library, where we would like to retrieve only the first 
paragraph of this paper.

The main complexity question of interest is the tradeoff between 
$m$ and $k$. With $k=\polylog(n)$ queries, 
the code length can be made polynomially small, even over 
the binary alphabet $\Sigma=\01$~\cite{bfls:checking}. 
However, for fixed $k$, the best upper bounds 
are superpolynomial. Except for the $k=2$ case with small 
alphabet $\Sigma$, no good lower bounds are known.  
Katz and Trevisan~\cite{katz&trevisan:ldc} showed superlinear
but at most quadratic lower bounds for constant $k$.
Goldreich et~al.~\cite{gkst:lowerpir} showed an exponential
lower bound for \emph{linear} codes with $k=2$ queries and constant alphabet,
and Kerenidis and de Wolf~\cite{kerenidis&wolf:qldcj} extended this
to \emph{all} codes, using techniques from quantum computing. 
For $\Sigma=\01^{\ell}$ they prove
$
m=2^{\Omega(n/2^{5\ell})}.
$
They also slightly improved the polynomial bounds 
of~\cite{katz&trevisan:ldc} for $k>2$.

Clearly the above lower bound becomes trivial if each position
of the codeword has $\ell\geq\log(n)/5$ bits.
In this paper we analyze the case where $\ell$ can be much larger,
but the decoder uses only $b$ bits out of the $\ell$ bits of a query answer.
The $b$ positions that he uses may depend on the
index $i$ he is interested in and on his randomness.
This setting is interesting because many existing constructions are 
of this form, for quite small $b$.
Goldreich et~al.~\cite{gkst:lowerpir} also analyzed this situation,
and showed the following lower bound for \emph{linear} codes:
$
m=2^{\Omega(n/\sum_{i=0}^b{\ell \choose i})}.
$
Here we prove a slightly weaker lower bound for all codes:
$
m=2^{\Omega(n/2^b\sum_{i=0}^b{\ell \choose i})}.
$
In particular, if $b=\ell$ (so the decoder can use all bits from the query answers)
we improve the bound from~\cite{kerenidis&wolf:qldcj} to
$
m=2^{\Omega(n/2^{2\ell})}.
$
We lose a factor of $2^b$ compared to Goldreich et~al. This factor
can be dispensed with if the decoder outputs the parity of
a subset of the bits he receives. All known LDCs are of this type.

Our proofs are completely different from the combinatorial approach of
Goldreich et al. Following~\cite{kerenidis&wolf:qldcj}, we proceed
in two steps: (1) we reduce the two classical queries to one 
\emph{quantum} query and (2) show a lower bound for the induced 
one-quantum-query-decodable code by deriving a \emph{random access code} from it.
The main novelty is a tight analysis of the following problem. 
Suppose we want to compute a Boolean
function $f(a_0,a_1)$ on $2b$ bits, given a quantum superposition
$\frac{1}{\sqrt{2}}(\ket{0,a_0}+\ket{1,a_1})$ of both halves of the input.
We show that \emph{any} Boolean $f$ can be computed with advantage
$1/2^{b+1}$ from this superposition, and that this is 
best-achievable for the parity function. This may be of 
independent interest.  In fact, Kerenidis~\cite{kerenidis:qcircuit} 
recently used it to exhibit an exponential quantum-classical 
separation in multiparty communication complexity, and in an
interesting new approach to improve depth lower bounds for \emph{classical} circuits.

\subsection{Private information retrieval}

There is a very close connection between LDCs and the setting of
\emph{private information retrieval}. In PIR, the user wants to retrieve some item
from a database without letting the database learn anything
about what item he asked for.
In the general model, the user retrieves the $i$th bit from an
$n$-bit database $x=x_1\ldots x_n$ that is replicated over
$k\geq 1$ non-communicating servers. He communicates
with each server without revealing any information about $i$ 
to individual servers, and at the end of the day learns $x_i$.
This is a natural cryptographic problem that has
applications in systems where privacy of the user is important,
for example databases providing medical information.
Much research has gone into optimizing the communication
complexity of one-round PIR schemes. Here the user sends a $t$-bit 
message (``query'') to each server, who responds with an 
$\ell$-bit message (``answer''), from which the user infers $x_i$.
A number of non-trivial upper bounds have been found
\cite{cgks:pir,ambainis:pir,beimel&ishai:unifiedpir,bikr:improvedpir}, 
but, as in the LDC case, the optimality of such schemes is wide open.
In fact, the best known constructions of LDCs with constant $k$ 
come from PIR schemes with $k$ servers.
Roughly speaking, concatenating the servers' answers 
to all possible queries gives a codeword $\cemph{C}(x)$ of length 
$m=k2^t$ over the alphabet $\Sigma = \01^{\ell}$ that is decodable with $k$ queries.
The privacy of the PIR scheme translates into the 
error-correcting property of the LDC: since many different sets of $k$ 
queries have to work for recovering $x_i$, we can afford some corrupted positions.
Conversely, we can turn a $k$-query LDC into a $k$-server
PIR scheme by asking one query to each server (so $t=\log m$). 
The privacy of the resulting PIR scheme follows from the fact 
that an LDC can be made to have a ``smoothness'' property, 
meaning that most positions are about equally likely to 
be queried, independent of $i$.

Here we restrict attention to 2 servers, which is probably the most interesting case.
The paper by Chor et~al.~\cite{cgks:pir} that introduced PIR,
gave a PIR scheme where both the queries 
to the servers and the answers from the servers have 
length $\Theta(n^{1/3})$ bits. Later constructions gave
alternative ways of achieving the same complexity,
but have not given asymptotic improvements for the 2-server case
(in contrast to the case of 3 or more servers~\cite{bikr:improvedpir}
and the case of 2 \emph{quantum} servers~\cite{kerenidis&wolf:qldcj}).
Though general lower bounds for 2-server PIRs still elude us,
reasonably good lower bounds can be proved for schemes that only use 
a small number $b$ of bits from each possibly much longer answer string. 
This $b$ is sometimes called the \emph{probe complexity} of the scheme.
As stated in~\cite{bik:generalpir}, small probe complexity is a desirable property 
of a PIR scheme for a number of reasons: the user needs 
less space; the schemes can be more easily applied recursively as 
in~\cite{bikr:improvedpir}; and such PIR schemes induce locally decodable codes where
the codelength $m$ is relatively small while the codeword entries are allowed 
to have many bits each, but the decoder needs only few bits from each 
codeword entry it read.

As was implicitly stated by Katz and Trevisan~\cite{katz&trevisan:ldc}
and formalized by Goldreich et al.~\cite{cgks:pir}, it is possible to translate
2-server PIRs to 2-query LDCs, where the property of only using $b$ bits
from each $\ell$-bit string carries over. 
Combining this lemma with our LDC lower bounds 
gives the following bound for 2-server PIRs
with $t$-bit queries, $\ell$-bit answers, and probe complexity $b$:
$
t=\Omega(n/2^b\sum_{i=0}^b{\ell \choose i}).
$
In particular, for fixed $b$ the overall communication is 
$C=2(t+\ell)=\Omega(n^{1/(b+1)})$.
This is tight for $b=1$ (we describe an $O(\sqrt{n})$ scheme 
in Section~\ref{ssecsquarescheme}) and close to optimal for $b=3$, 
since a small variation of the Chor et al.~scheme achieves 
$C=O(n^{1/3})$ using only 3 bits from each answer
\footnote{A polynomial-based
$O(n^{1/3})$-scheme from~\cite{beimel&ishai:unifiedpir} 
does not have this ``small $b$''-property.}, while our bound is $\Omega(n^{1/4})$.
Similar results were established for \emph{linear} PIR schemes
by Goldreich et al., but our results apply to \emph{all} PIR schemes.
They imply that in improved 2-server PIR schemes, 
the user needs to use more bits from the servers' answers. 
For general schemes, where $b = \ell$, we obtain
$
t=\Omega(n/2^{2\ell}).
$
This improves the $n/2^{5\ell}$ bound from~\cite{kerenidis&wolf:qldcj}.
It implies a lower bound of $5\log n$ on the total communication
$C=2(t+\ell)$. This is incredibly weak, but without any assumptions
on how the user handles the answers, and still improves
what was known~\cite{mann:pir,kerenidis&wolf:qldcj}.

\section{Preliminaries}\label{secprelim}

We use $a_{|S}$ to denote the string $a$ restricted to a 
set of bits $S\subseteq[n]=\{1,\ldots,n\}$, e.g., $11001_{|\{1,4,5\}}=101$.
We identify a set $S\subseteq[n]$ with $n$-bit string $S=S_1\ldots S_n$, 
where $i \in S$ if and only if the $i$th bit $S_i = 1$.
We use $e_i$ for the $n$-bit string corresponding to the singleton set $S=\{i\}$.
If $y\in\Sigma^m$ where $\Sigma =\01^\ell$, 
then $y_j \in \Sigma$ denotes its $j$th entry, 
and $y_{j,i}$ with $i\in[\ell]$ is the $i$th bit of $y_j$.
We assume general familiarity with the quantum model~\cite{nielsen&chuang:qc}.
Our proofs depend heavily on the notion of a quantum query.
We consider queries with $\ell$-bit answers, where $\ell \geq 1$.
For $\Sigma=\01^\ell$, a quantum query to a string $y \in \Sigma^m$
is the unitary map
$
\ket{j}\ket{z} \mapsto \ket{j}\ket{z \oplus y_j},
$ 
where $j \in [m]$, $z \in \01^{\ell}$ is called the target register, and
$z \oplus y_j$ is the string resulting from the xor
of the individual bits of $z$ and $y_j$, i.e.
$z \oplus y_j = (z_1 \oplus y_{j,1})\ldots (z_{\ell} \oplus y_{j,\ell})$.
It is convenient to get the query result in the phase of the quantum state.
To this end, define
$
\ket{z_T} = \frac{1}{\sqrt{2^{\ell}}}\bigotimes_{i = 1}^{\ell}(\ket{0} + (-1)^{T_i} \ket{1})
$
where $T_i$ is the $i$th bit of the $\ell$-bit string $T$. 
Since $\ket{0\oplus y_{j,i}}+(-1)^{T_i}\ket{1\oplus y_{j,i}}=
(-1)^{T_i\cdot y_{j,i}}(\ket{0}+(-1)^{T_i}\ket{1})$,
a query maps
$
\ket{j}\ket{z_T} \mapsto \ket{j}(-1)^{T \cdot y_j}\ket{z_T}.
$

A locally decodable code is an 
error-correcting code that allows efficient decoding of individual bits.

\begin{definition}
$\cemph{C}:\01^n \rightarrow \Sigma^m$ is a $(k,\delta,\eps)$-\emph{locally decodable code (LDC)}, if
there exists a classical randomized decoding algorithm $A$ with input $i\in[n]$ 
and oracle access to a string $y\in\Sigma^m$ such that
\begin{enumerate}
\item $A$ makes $k$ distinct queries $j_1,\ldots,j_k$ to $y$, 
non-adaptively, gets query answers $a_1=y_{j_1},\ldots,a_k=y_{j_k}$ and outputs 
a bit $f(a_1,\dots,a_k)$, where $f$ depends on $i$ and $A$'s randomness.
\item For every $x \in \01^n$, $i\in[n]$ and $y \in \Sigma^m$ 
with Hamming distance $d(y,\cemph{C}(x)) \leq \delta m$ we have
$\Pr[f(a_1,\ldots,a_k) = x_i] \geq 1/2 + \eps$.
\end{enumerate}
Here probabilities are taken over $A$'s internal randomness.
For $\Sigma=\01^\ell$, we say the LDC \emph{uses $b$ bits},
if $A$ only uses $b$ predetermined bits of each query answer: 
it outputs $f(a_{1|S_1},\ldots,a_{k|S_k})$ where the sets $S_1,\ldots,S_k$ 
are of size $b$ each and are determined by $i$ and $A$'s randomness.
\end{definition}


In our arguments we will use \emph{smooth} codes. 
These are codes where the decoding algorithm spreads its queries ``smoothly'' across the
codeword, meaning it queries no code location too frequently.

\begin{definition}
$\cemph{C}:\01^n \rightarrow \Sigma^m$ is a $(k,c,\eps)$-\emph{smooth code (SC)} if
there is a randomized algorithm $A$ 
with input $i\in[n]$ and oracle access to $\cemph{C}(x)$ s.t.
\begin{enumerate}
\item $A$ makes $k$ distinct queries $j_1,\ldots,j_k$ to $\cemph{C}(x)$, 
non-adaptively, gets query answers 
$a_1=\cemph{C}(x)_{j_1},\ldots,a_k=\cemph{C}(x)_{j_k}$ and outputs 
a bit $f(a_1,\dots,a_k)$, where $f$ depends on $i$ and $A$'s randomness.
\item For every $x \in \01^n$ and $i \in [n]$ we have
$\Pr[f(a_1,\ldots,a_k) = x_i] \geq 1/2 + \eps.$
\item For every $x\in\01^n$, $i \in [n]$ and $j \in [m]$, 
$\Pr[A\mbox{ queries }j]\leq c/m$.
\end{enumerate}
The smooth code \emph{uses $b$ bits}, if $A$ only uses $b$ predetermined bits of each answer.
\end{definition}

Note that the decoder of smooth codes deals only with valid codewords $\cemph{C}(x)$. 
The decoding algorithm of an LDC on the other hand can deal with 
corrupted codewords $y$ that are still sufficiently close to the original.
Katz and Trevisan~\cite[Theorem 1]{katz&trevisan:ldc} 
showed that LDCs and smooth codes are closely related:

\begin{theorem}[Katz \&\ Trevisan]\label{trevisan_smooth}
If $\cemph{C}: \01^n \rightarrow \Sigma^m$ is a $(k,\delta,\eps)$-LDC,
then $\cemph{C}$ is also a $(k,k/\delta,\eps)$-smooth code (the property of using $b$ bits carries over).
\end{theorem}

The following definition of a one-query \emph{quantum} smooth code is rather ad hoc and
not the most general possible, but sufficient for our purposes.

\begin{definition}
$\cemph{C}:\01^n \rightarrow \Sigma^m$ is a $(1,c,\eps)$-\emph{quantum smooth code (QSC)},
if there is a quantum algorithm $A$ 
with input $i\in[n]$ and oracle access to $\cemph{C}(x)$ s.t.
\begin{enumerate}
\item $A$ probabilistically picks a string $r$,  
makes a query of the form
$$
\ket{Q_{ir}} =\frac{1}{\sqrt{2}}\left(
\ket{j_{1r}}\frac{1}{\sqrt{2^b}}\sum_{T\subseteq S_{1r}}\ket{z_T}+
\ket{j_{2r}}\frac{1}{\sqrt{2^b}}\sum_{T\subseteq S_{2r}}\ket{z_T}
\right)
$$
and returns the outcome of some measurement on the resulting state.
\item For every $x \in \01^n$ and  $i \in [n]$ we have $\Pr[A\mbox{ outputs }x_i] \geq 1/2 + \eps$.
\item For every $x,i,j$, 
$\Pr[A\mbox{ queries }j\mbox{ with non-zero amplitude}]\leq c/m$.
\end{enumerate}
The QSC \emph{uses $b$ bits}, if the sets $S_{1r}, S_{2r}$ have size $b$.
\end{definition}


PIR allows a user to obtain the $i$th bit from an $n$-bit
database $x$, replicated over $k\geq 1$ servers, without revealing anything
about $i$ to individual servers.

\begin{definition}
A one-round, $(1-\eta)$-secure, $k$-server \emph{private information retrieval (PIR)} 
scheme for a database $x \in \01^n$ with 
recovery probability $1/2 + \eps$, query size $t$, and answer size $\ell$, consists of 
a randomized algorithm (user) and $k$ deterministic algorithms
$S_1,\ldots,S_k$ (servers), such that
\begin{enumerate}
\item On input $i \in [n]$, the user produces $k$ $t$-bit queries $q_1,\ldots,q_k$
and sends these to the respective servers. The $j$th server returns $\ell$-bit
string $a_j = S_j(x,q_j)$. The user outputs a bit $f(a_1,\ldots,a_k)$ ($f$
depends on $i$ and his randomness).
\item For every $x \in \01^n$ and $i \in [n]$ we have
$\Pr[f(a_1,\ldots,a_k) = x_i] \geq 1/2 + \eps.$
\item For all $x\in\01^n$, $j\in[k]$, and any two indices $i_1,i_2\in[n]$,
the two distributions on $q_j$ (over the user's randomness)
induced by $i_1$ and $i_2$ are $\eta$-close in total variation distance.
\end{enumerate}
The scheme \emph{uses $b$ bits} if the user only uses $b$ predetermined 
bits from each $a_i$.
The scheme is called \emph{linear}, if for every $j$ and $q_j$ the $j$th server's 
answer $S_j(x,q_j)$ is a linear combination (over $GF(2)$) of the bits of $x$.
\end{definition}

If $\eta = 0$, then the server gets no information
at all about $i$. All known non-trivial PIR schemes have $\eta=0$, perfect
recovery ($\eps = 1/2$), and one round of communication.
We give two well-known 2-server examples from~\cite{cgks:pir}.\\[3mm]
{\bf Square scheme.}\label{ssecsquarescheme}
Arrange $x = x_1\ldots x_n$ in a $\sqrt{n}\times\sqrt{n}$ square,
$$
x=\left(
\begin{array}{ccccc}
x_1           & x_2 & &\cdots& x_{\sqrt{n}}\\
x_{\sqrt{n}+1} & \ddots & & & x_{2\sqrt{n}} \\
\vdots        & & x_i & & \vdots \\
\vdots        & \cdots & \cdots & \cdots & x_n
\end{array}
\right)
$$
then index $i$ is given by two coordinates $(i_1,i_2)$.
The user picks a random string $A\in\01^{\sqrt{n}}$, and sends
$\sqrt{n}$-bit queries $q_1 = A$ and $q_2 = A \oplus e_{i_1}$ to the 
servers.  The first returns $\sqrt{n}$-bit 
answer $a_1 = q_1\cdot C_1,\ldots,q_1\cdot C_{\sqrt{n}}$, where $q_1\cdot C_c$
denotes the inner product mod 2 of $q_1$ with the $c$th column of $x$.
The second server sends $a_2$ analogously.
The user selects the bit $q_1\cdot C_{i_2}$ from $a_1$ and $q_2\cdot C_{i_2}$ 
from $a_2$ and computes
$(A \cdot C_{i_2}) \oplus ((A \oplus e_{i_1}) \cdot C_{i_2})
=e_{i_1} \cdot C_{i_2} = x_i$.
Here $t =\ell= \sqrt{n}$ and $b=1$.\\[2mm]
{\bf Cube scheme.}\label{sseccubescheme}
A more efficient scheme arranges $x$ in a cube, so $i=(i_1,i_2,i_3)$. 
The user picks 3 random strings $T_1$, $T_2$, $T_3$ of $n^{1/3}$ bits each, and sends queries
$q_1 = T_1, T_2, T_3$ and $q_2 = (T_1\oplus e_{i_1}),(T_2\oplus e_{i_2}),(T_3\oplus e_{i_3})$.
The first server computes the bit
$
a=b_{T_1T_2T_3} = \bigoplus_{j_1\in T_1,j_2\in T_2,j_3\in T_3} x_{j_1,j_2,j_3}.
$
Its answer $a_1$ is the $n^{1/3}$ bits $b_{T'_1T_2T_3}\oplus a$
for all $T'_1$ differing from $T_1$ in exactly one place, 
and similarly all $b_{T_1T'_2T_3}\oplus b$ and $b_{T_1T_2T'_3}\oplus a$. 
The second server does the same with its query $q_2$.
The user now selects those 3 bits of each answer that correspond
to $T'_1=T_1\oplus e_{i_1}$, $T'_2=T_2\oplus e_{i_2}$, $T'_3=T_3\oplus e_{i_3}$ 
respectively, and xors those 6 bits. Since every other 
$x_{j_1,j_2,j_3}$ occurs exactly twice in that sum, 
what is left is $x_{i_1,i_2,i_3} = x_i$.
Here $t,\ell= O(n^{1/3})$ and $b=3$.
%
%
\section{Computing $f(a_0,a_1)$ from Superposed Input}

\subsection{Upper bound}

To prove the lower bound on LDCs and PIRs, we first construct
the following quantum tool. 
Consider a state $\ket{\Psi_{a_0a_1}}=\frac{1}{\sqrt{2}}(\ket{0,a_0}+\ket{1,a_1})$
with $a_0,a_1$ both $b$-bit strings.
We show that we can compute any Boolean function $f(a_0,a_1)$ with bias $1/2^{b+1}$ 
given one copy of this state. 
After that we show that bias is optimal if $f$ is the $2b$-bit parity function.
The key to the algorithm is the following:


\begin{lemma}\label{lemgammaunitary}
For every $f:\01^{2b} \rightarrow \01$ there exist
non-normalized states $\ket{\varphi_{a}}$ such that
$
U: \ket{a}\ket{0} \rightarrow\frac{1}{2^b} \sum_{w \in \01^b} (-1)^{f(w,a)}\ket{w}\ket{0} + 
\ket{\varphi_{a}}\ket{1}
$ 
is unitary.
\end{lemma}

\begin{proof}
Let $\ket{\psi_{a}} = (1/2^b) \sum_{w\in \01^b} (-1)^{f(w,a)} \ket{w}\ket{0} + \ket{\varphi_{a}}\ket{1}$. 
It is easy to see that $U$ can be extended to be unitary if and only if $\inp{\psi_{a}}{\psi_{a'}} = \delta_{aa'}$
for all $a,a'$. We will choose $\ket{\varphi_{a}}$ to achieve this.
First, since $\inp{w}{w'} = \delta_{ww'}$ and $\inp{w,0}{\varphi_{a},1} = 0$:
$$
\inp{\psi_{a}}{\psi_{a'}} = \frac{1}{2^{2b}} \sum_{w\in \01^b} (-1)^{f(w,a) + f(w,a')}  +
 \inp{\varphi_{a}}{\varphi_{a'}}.
$$
Let $C$ be the $2^b \times 2^b$ matrix with entries
$C_{aa'} =(1/2^{2b}) \sum_{w\in \01^b} (-1)^{f(w,a) + f(w,a')}$ 
where the indices $a$ and $a'$ are $b$-bit strings.  
{}From the definition of $C_{aa'}$ we have $|C_{aa'}|\leq 1/2^b$.
By~\cite[Corollary~6.1.5]{horn&johnson:ma}, the largest eigenvalue is
$$
\lambda_{max}(C) \leq \min \left\{ \max_a \sum_{a'\in\01^b} |C_{aa'}|, \max_{a'} \sum_{a\in\01^b} |C_{aa'}|\right\} \leq 
\sum_{a \in\01^b} \frac{1}{2^b} = 1.
$$
However, $\lambda_{max}(C) \leq 1$ implies that 
$I - C$  is positive semidefinite and hence,
by~\cite[Corollary~7.2.11]{horn&johnson:ma}, $I - C = A^{\dagger}A$ for some 
matrix $A$. Now define $\ket{\varphi_{a}}$ to be the $a$th column of $A$.
Since the matrix $C + A^{\dagger}A = I$ is composed of all inner products 
$\inp{\psi_{a}}{\psi_{a'}}$, we have $\inp{\psi_{a}}{\psi_{a'}}=\delta_{aa'}$ and 
it follows that $U$ is unitary.
\end{proof}

Using these observations, we can now prove the following theorem.

\begin{theorem}\label{fsup}
Suppose $f: \01^{2b} \rightarrow \01$ is a Boolean function. 
There exists a quantum algorithm to compute $f(a_0,a_1)$ 
with success probability exactly $1/2 + 1/2^{b+1}$ using one copy of 
$\ket{\Psi_{a_0a_1}}=\frac{1}{\sqrt{2}}(\ket{0,a_0} + \ket{1,a_1})$, with $a_0,a_1 \in \01^b$.
\end{theorem}

\begin{proof}
First we extend the state $\ket{\Psi_{a_0a_1}}$ by a $\ket{0}$-qubit.
Let $U$ be as in Lemma~\ref{lemgammaunitary}.
Applying the unitary transform $\outp{0}{0} \otimes  I^{\otimes b+1} + \outp{1}{1} \otimes U$ 
to $\ket{\Psi_{a_0a_1}}\ket{0}$ gives
$$
\frac{1}{\sqrt{2}}\left(
\ket{0}\ket{a_0}\ket{0} + 
\ket{1}\left(\frac{1}{2^b} \sum_{w\in\01^b}(-1)^{f(w,a_1)} \ket{w}\ket{0} + \ket{\varphi_{a_1}}\ket{1}\right)
\right).
$$
Define $\ket{\Gamma} = \ket{a_0}\ket{0}$ and $\ket{\Lambda} = 
\frac{1}{2^b} \sum_w {(-1)^{f(w,a_1)} \ket{w}}\ket{0} + \ket{\varphi_{a_1}}\ket{1}$.
Then $\inp{\Gamma}{\Lambda} = \frac{1}{2^b}(-1)^{f(a_0,a_1)}$ and
the above state is
$
\frac{1}{\sqrt{2}}(\ket{0}\ket{\Gamma} + \ket{1}\ket{\Lambda}).
$
We apply a Hadamard transform to the first qubit to get
$
\frac{1}{2} \left(\ket{0}(\ket{\Gamma} + \ket{\Lambda}) + \ket{1}(\ket{\Gamma} - \ket{\Lambda})\right).
$
The probability that a measurement of the first qubit yields a 0 is
$
\frac{1}{4} \inp{\Gamma + \Lambda}{\Gamma + \Lambda} = 
\frac{1}{2} + \frac{1}{2} \inp{\Gamma}{\Lambda} = 
\frac{1}{2}+\frac{(-1)^{f(a_0,a_1)}}{2^{b+1}}.
$
Thus by measuring the first qubit we obtain $f(a_0,a_1)$ with bias $1/2^{b+1}$. 
\end{proof}

\subsection{Lower bound}

To prove that this algorithm is optimal for the parity function, 
we need to consider how well we can distinguish two density matrices
$\rho_0$ and $\rho_1$, i.e., given an unknown state
determine whether it is $\rho_0$ or $\rho_1$. Let $\norm{A}_{tr}$ 
denote the trace norm of matrix $A$, which equals the sum of its singular values.

\begin{lemma}\label{distinguish}
Two density matrices $\rho_0$ and $\rho_1$ cannot be distinguished with
probability better than $1/2+\norm{\rho_0-\rho_1}_{tr}/4$.
\end{lemma}

\begin{proof}
The most general way of distinguishing $\rho_0$ and $\rho_1$ is
a POVM~\cite{nielsen&chuang:qc} with two operators $E_0$ and $E_1$, such that 
$p_0 = tr(\rho_0 E_0) \geq 1/2 + \eps$ and 
$q_0 = tr(\rho_1 E_0) \leq 1/2 - \eps$.
Then $|p_0 - q_0| \geq 2 \eps$ and likewise, $|p_1 - q_1| \geq 2 \eps$, for
similarly defined $p_1$ and $q_1$.
By ~\cite[Theorem 9.1]{nielsen&chuang:qc}, 
$\norm{\rho_0-\rho_1}_{tr} = \max_{\{E_0,E_1\}}(|p_0 - q_0| + |p_1 - q_1|)$ 
and thus $\norm{\rho_0 - \rho_1}_{tr} \geq 4 \eps$.
Hence $\eps\leq\norm{\rho_0 - \rho_1}_{tr}/4$.
\end{proof}

\begin{theorem}
Suppose that $f$ is the parity of $a_0a_1$.
Then any quantum algorithm for computing $f$ from one copy
of $\ket{\Psi_{a_0a_1}}$ has success probability $\leq 1/2+1/2^{b+1}$.
\end{theorem}
\begin{proof}
Define $\rho_0$ and $\rho_1$ by
$
\rho_c=\frac{1}{2^{2b-1}}\sum_{a_0a_1\in f^{-1}(c)}\ketbra{\Psi_{a_0a_1}}{\Psi_{a_0a_1}},
$
with $c \in \01$. 
A quantum algorithm that computes parity of $a_0a_1$ with probability $1/2+\eps$
can be used to distinguish $\rho_0$ and $\rho_1$. Hence by
Lemma~\ref{distinguish}: $\eps \leq \norm{\rho_0 - \rho_1}_{tr}/4$. 
Let $A=\rho_0-\rho_1$.
It is easy to see that the $\ketbra{0,a_0}{0,a_0}$-entries
are the same in $\rho_0$ and in $\rho_1$, so these entries are 0 in $A$.
Similarly, the $\ketbra{1,a_1}{1,a_1}$-entries in $A$ are 0.
In the off-diagonal blocks, the $\ketbra{0,a_0}{1,a_1}$-entry
of $A$ is $(-1)^{|a_0| + |a_1|}/2^{2b}$.
For $\ket{\phi} = \frac{1}{\sqrt{2^b}} \sum_{w \in \01^{b}} (-1)^{|w|} \ket{w}$ we have
$
\outp{\phi}{\phi} = \frac{1}{2^b} \sum_{a_0,a_1} (-1)^{|a_0| + |a_1|} \outp{a_0}{a_1}
$ and 
$
A = \frac{1}{2^b} (\outp{0,\phi}{1,\phi} + \outp{1,\phi}{0,\phi}).
$
Let $U$ and $V$ be unitary transforms such 
that $U\ket{0,\phi} = \ket{0,0^b}$, $U\ket{1,\phi} = \ket{1,0^b}$ and
$V\ket{0,\phi} = \ket{1,0^b}$, $V\ket{1,\phi} = \ket{0,0^b}$,
then 
$
U A V^{\dagger} = \frac{1}{2^b} (U \outp{0,\phi}{1,\phi} V^{\dagger}
+ U \outp{1,\phi}{0,\phi} V^{\dagger}) = \frac{1}{2^b} (\outp{0,0^b}{0,0^b} + \outp{1,0^b}{1,0^b}).
$
The two nonzero singular values of $U A V^{\dagger}$ are both $1/2^b$, hence
$
\norm{\rho_0-\rho_1}_{tr}=\norm{A}_{tr}= \norm{U A V^{\dagger}}_{tr} = 2/2^{b}.
$
Therefore $\eps\leq \norm{\rho_0-\rho_1}_{tr}/4=1/2^{b+1}$.
\end{proof}

%
%

\section{Lower Bounds for LDCs that Use Few Bits}

We now make use of the technique developed above to prove new lower 
bounds for 2-query LDCs over non-binary alphabets. First we 
construct a 1-query quantum smooth code (QSC) from a 2-query smooth 
code (SC), and then prove lower bounds for QSCs.
In the sequel, we will index the two queries by 0 and 1
instead of 1 and 2, to conform to the two basis states 
$\ket{0}$ and $\ket{1}$ of a qubit.

\subsection{Constructing a 1-query QSC from a 2-query SC}

\begin{theorem}\label{lqdc}
If $\cemph{C}: \01^n \rightarrow(\01^{\ell})^m$ is a $(2,c,\eps)$-smooth code
that uses $b$ bits, then $\cemph{C}$ is a $(1,c,\eps/2^b)$-quantum smooth 
code that uses $b$ bits. 
\end{theorem}

\begin{proof}
Fix index $i\in[n]$ and encoding $y = \cemph{C}(x)$. 
The 1-query quantum decoder will pick a random string $r$ with the same 
probability as the 2-query classical decoder.
This $r$ determines two indices $j_0,j_1\in [m]$, two $b$-element sets 
$S_0,S_1\subseteq[\ell]$, and a function $f: \01^{2b} \rightarrow \01$ such that
$
\Pr[f(y_{j_0|S_0},y_{j_1|S_1}) = x_i] = p \geq \frac{1}{2} + \eps,
$
where the probability is taken over the decoder's randomness. 
Assume for simplicity that $j_0=0$ and $j_1=1$, and define 
$a_0=y_{j_0|S_0}$ and $a_1=y_{j_1|S_1}$.
We now construct a 1-query quantum decoder that outputs $f(a_0,a_1)$ 
with probability $1/2 + 1/2^{b+1}$, as follows. 
The result of a quantum query to $j_0$ and $j_1$ is
$$
\frac{1}{\sqrt{2}}
\left(\underbrace{\ket{0}}_{j_0}\frac{1}{\sqrt{2^b}}\sum_{T \subseteq S_0}(-1)^{a_0 \cdot T}\ket{z_T} +
\underbrace{\ket{1}}_{j_1}\frac{1}{\sqrt{2^b}}\sum_{T \subseteq S_1}(-1)^{a_1 \cdot T}\ket{z_T}\right).
$$
Note that we write $a_0 \cdot T$ instead of $y_{j_0} \cdot T$, since $T \subseteq S_0$ and
therefore the inner product will be the same.
We can unitarily transform this to
$
\frac{1}{\sqrt{2}}(\ket{0}\ket{a_0} + \ket{1}\ket{a_1}).
$
By Theorem~\ref{fsup}, we can compute an output bit $o$ from this such that
$\Pr[o=f(a_0,a_1)]=1/2 + 1/2^{b+1}$.                                          
The probability of success is then given by
$\Pr[o = x_i] = \Pr[o = f(a_0,a_1)]\Pr[x_i = f(a_0,a_1)] + 
                 \Pr[o \neq f(a_0,a_1)]\Pr[x_i \neq f(a_0,a_1)] 
  = (1/2 + 1/2^{b+1})p + (1/2 - 1/2^{b+1})(1-p) \geq 1/2 + \eps/2^b$.
Since no $j$ is queried with probability more than $c/m$
by the classical decoder, the same is true for the quantum decoder.
\end{proof}
\subsection{Improved lower bounds for 2-query LDCs over an $\ell$-bit alphabet}
Our lower bound for 2-query LDCs uses the following notion,
due to~\cite{ambainis:rac}. 

\begin{definition}
A \emph{quantum random access code} is a mapping $x\mapsto\rho_x$ 
of the $n$-bit strings $x$ into $m$-qubit states $\rho_x$, 
such that any bit $x_i$ can be recovered with some probability 
$p\geq 1/2+\eps$ from $\rho_x$
\end{definition}
Note that we need not be able to recover \emph{all} $x_i$'s simultaneously
from $\rho_x$, just any one $x_i$ of our choice.
Nayak~\cite{nayak:qfa} proved a tight bound on $m$:

\begin{theorem}[Nayak]\label{thnayak}
Every quantum random access code has $m \geq (1 - H(p))n$.
\end{theorem}

The main idea of our proof is to show how the following state $\ket{U(x)}$ 
induces a quantum random access code. 
For $u = \sum_{i=0}^b {\ell \choose i}$ define the pure states
$$
\ket{U(x)_j}=\frac{1}{\sqrt{u}}\sum_{|T|\leq b}(-1)^{T \cdot \cemph{C}(x)_j} \ket{z_T}
\mbox{ and }
\ket{U(x)} = \frac{1}{\sqrt{m}}\sum_{j=1}^m \ket{j}\ket{U(x)_j}.
$$

\begin{lemma}\label{raccess}
Suppose $\cemph{C}:\01^n \rightarrow (\01^{\ell})^m$ is a $(1,c,\eps)$-quantum smooth
code that uses $b$ bits. Then given one copy of $\ket{U(x)}$, there is a quantum algorithm that 
outputs `fail' with probability $1-2^{b+1}/(cu)$ with $u = \sum_{i=0}^b{\ell \choose i}$, 
but if it succeeds it outputs $x_i$ with probability at least $1/2+\eps$.
\end{lemma}

\begin{proof}
Let us fix $i\in[n]$. Suppose the quantum decoder of $\cemph{C}$
makes query $\ket{Q_{ir}}$ to indices $j_{0r}$ and $j_{1r}$ 
with probability $p_r$. Consider 
the following state
$$
\ket{V_i(x)}=\sum_r\sqrt{p_r}\ket{r}\frac{1}{\sqrt{2}}
\left(\ket{j_{0r}}\ket{U(x)_{j_{0r}}}+\ket{j_{1r}}\ket{U(x)_{j_{1r}}}\right).
$$
We first show how to obtain $\ket{V_i(x)}$ from $\ket{U(x)}$ with
some probability. 
Rewrite
$$
\ket{V_i(x)}=\sum_{j=1}^m \alpha_j \ket{\phi_j}\ket{j}\ket{U(x)_j},
$$
where the $\alpha_j$ are nonnegative reals, and $\alpha^2_j\leq c/(2m)$
because $C$ is a QSC (the $1/2$ comes from the amplitude $1/\sqrt{2}$). 
Using the unitary map $\ket{0}\ket{j}\mapsto\ket{\phi_j}\ket{j}$,
we can obtain $\ket{V_i(x)}$ from the state
$
\ket{V'_i(x)}=\sum_{j=1}^m \alpha_j\ket{j}\ket{U(x)_j}.
$
We thus have to show that we can obtain $\ket{V'_i(x)}$ from $\ket{U(x)}$.
Define operator 
$
M = \sqrt{2m/c} \sum_{j=1}^m \alpha_j \outp{j}{j} \otimes I
$
and consider a POVM with operators $M^\dagger M$ and $I-M^\dagger M$.
These operators are positive because $\alpha_j^2\leq c/2m$.
Up to normalization, $M\ket{U(x)}=\ket{V'_i(x)}$.
The probability that the measurement succeeds (takes us from
$\ket{U(x)}$ to $\ket{V'_i(x)}$) is
$
\bra{U(x)}M^\dagger M\ket{U(x)} = \frac{2m}{c}\bra{U(x)}\left(\sum_j \alpha_j^2 \outp{j}{j} \otimes I\right)\ket{U(x)}
= \frac{2}{c}\sum_j \alpha_j^2
= \frac{2}{c}.
$
Now given $\ket{V_i(x)}$ we can measure $r$, and then project the last register
onto the sets $S_{0r}$ and $S_{1r}$ that we need for $\ket{Q_{ir}}$,
by means of the measurement operator
$
\outp{j_{0r}}{j_{0r}}\otimes\sum_{T\subseteq S_{0r}}\outp{T}{T}+
\outp{j_{1r}}{j_{1r}}\otimes\sum_{T\subseteq S_{1r}}\outp{T}{T}.
$
This measurement succeeds with probability $2^b/u$,
but if it succeeds we have the state corresponding to the answer to
query $\ket{Q_{ir}}$, from which we can predict $x_i$.
Thus, 
we succeed with probability $(2^b/u)\cdot(2/c)$,
and \emph{if} we succeed, we output $x_i$ with probability $1/2+\eps$.
\end{proof}
We can avoid failures by taking many copies of $\ket{U(x)}$:
\begin{lemma}\label{raccessCol}
If $\cemph{C}:\01^n \rightarrow (\01^{\ell})^m$ is a $(1,c,\eps)$-quantum smooth 
code, then $\ket{W(x)} = \ket{U(x)}^{\otimes cu/2^{b+1}}$  
is a $cu(\log(m) + \log(u))/2^{b+1}$-qubit random access code for $x$ with 
recovery probability $1/2 + \eps/2$ where $u=\sum_{i=0}^{b}{\ell \choose i}$.
\end{lemma}
\begin{proof}
We do the experiment of the previous lemma on each copy of $\ket{U(x)}$
independently. The probability that all experiments fail simultaneously 
is $(1-2^{b+1}/(cu))^{cu/2^{b+1}}\leq 1/2$. 
In that case we output a fair coin flip. If at least one experiment succeeds,
we can predict $x_i$ with probability $1/2+\eps$. 
This gives overall success probability at least
$1/2(1/2 + \eps) + (1/2)^2 = 1/2 + \eps/2$.
\end{proof}
The lower bound for 
2-query SCs and LDCs over non-binary alphabets is then:

\begin{theorem}\label{mtwoquery}
If $\cemph{C}: \01^n \rightarrow \Sigma^m = (\01^{\ell})^m$ is a
$(2,c,\eps)$-smooth code where the decoder uses only $b$ bits of each answer, 
then 
$
m \geq 2^{dn - \log(u)}
$
for $d = (1 - H(1/2 + \eps/2^{b+1}))2^{b+1}/(cu)=\Theta(\eps^2/(2^b c u))$ and
$u = \sum_{i=0}^b{\ell \choose i}$.
Hence $m=2^{\Omega(\eps^2n/(2^{2\ell}c))}$ if $b=\ell$.
\end{theorem}
\begin{proof}
Theorem~\ref{lqdc} implies that $\cemph{C}$ is a $(1,c,\eps/2^b)$-quantum smooth code.
Lemma~\ref{raccessCol} gives us a random access code of 
$cu(\log(m) + \log(u))/2^{b+1}$ qubits with recovery probability $p = 1/2 + \eps/2^{b+1}$. 
Finally, the random access code lower bound, Theorem~\ref{thnayak},
implies $cu(\log(m) + \log(u))/2^{b+1}\geq(1 - H(p))n$.
Rearranging and using that $1-H(1/2+\eta)=\Theta(\eta^2)$ gives the result.
\end{proof}
Since a $(2,\delta,\eps)$-LDC is a $(2,2/\delta,\eps)$-smooth code 
(Theorem~\ref{trevisan_smooth}), we obtain:

\begin{corollary}
If $\cemph{C}: \01^n \rightarrow \Sigma^m = (\01^{\ell})^m$ is a
$(2,\delta,\eps)$-locally decodable code, then 
$
m \geq 2^{dn - \log(u)}
$
for $d = (1 - H(1/2 + \eps/2^{b+1}))\delta 2^b/u=\Theta(\delta\eps^2/(2^b u))$ and
$u = \sum_{i=0}^b{\ell \choose i}$.
Hence $m=2^{\Omega(\delta\eps^2 n/2^{2\ell})}$ if $b=\ell$.
\end{corollary}
In all known non-trivial constructions of LDCs and SCs,
the decoder outputs the parity of the bits that he is interested in.
Then, we can prove:

\begin{theorem}\label{mtwoqueryparity}
If $\cemph{C}: \01^n \rightarrow \Sigma^m = (\01^{\ell})^m$ is a $(2,c,\eps)$-smooth 
code where the decoder outputs $f(g(a_{0|S_0}),g(a_{1|S_1}))$, 
with $f,g:\01^2\rightarrow\01$ fixed functions, then
$
m \geq 2^{dn - \log(\ell')}
$
for $d = \Omega(\eps^2/(c\ell'))$ and $\ell' = {\ell \choose b}$.
\end{theorem}

\begin{proof}
Transform $\cemph{C}$ into a smooth code 
$\cemph{C'}: \01^n \rightarrow (\01^{\ell'})^m$ with 
$\ell' = {\ell \choose b}$ by defining $\cemph{C'}(x)_j$ to be the value of $g$ on all 
${\ell \choose b}$ possible $b$-subsets of the original $\ell$ bits of $\cemph{C}(x)_j$.
We need only 1 bit of each $\cemph{C'}(x)_j$, and can apply
Theorem~\ref{mtwoquery}.
\end{proof}

%
%

\section{Lower Bounds for Private Information Retrieval}

\subsection{Lower bounds for 2-server PIRs that use few bits}

Here we derive improved lower bounds for 2-server PIRs 
from our LDC bounds.
We use the following~\cite[Lemma 7.1]{gkst:lowerpir}
to translate PIR schemes to smooth codes: 

\begin{lemma}[GKST]\label{pirldc}
Suppose there is a one-round, $(1-\eta)$-secure PIR scheme with two servers, database size
$n$, query size $t$, answer size $\ell$, and recovery probability at least $1/2 + \eps$.  
Then there is a $(2,3,\eps-\eta)$-smooth code 
$\cemph{C}: \01^n \rightarrow (\01^{\ell})^m$, where $m \leq 6 \cdot 2^t$.
If the PIR scheme uses only $b$ bits of each server answer,
then the resulting smooth code uses only $b$ bits of each query answer.
\end{lemma}
We now combine this with Theorem~\ref{mtwoquery}
to slightly improve the lower bound
given in~\cite{kerenidis&wolf:qldcj} and to extend it to the case where we 
only use $b$ bits of each server reply.

\begin{theorem}\label{pirlarger}
A classical 2-server $(1-\eta)$-secure PIR scheme with $t$-bit queries, 
$\ell$-bit answers that uses $b$ bits and has recovery probability $1/2 + \eps$ satisfies
$
t = \Omega\left(\frac{n(\eps-\eta)^2}{2^{b}u}\right)
$
with $u = \sum_{i = 0}^b {\ell \choose i}$. In particular, if $b=\ell$, 
then $t=\Omega(n(\eps-\eta)^2/2^{2\ell})$.
\end{theorem}

\begin{proof}
Using Lemma~\ref{pirldc} we turn the PIR scheme into a $(2,3,\eps-\eta)$-smooth code 
$\cemph{C}: \01^n \rightarrow (\01^{\ell})^m$ that uses $b$ bits
of $\ell$ where $m \leq 6 \cdot 2^t$. 
{}From Theorem~\ref{mtwoquery} we have $m \geq 2^{dn - \log(u)}$ with 
$d = \Theta((\eps-\eta)^2/(2^bu))$. 
\end{proof}
If $b$ is fixed, $\eps=1/2$ and $\eta=0$, this bound simplifies 
to $t=\Omega(n/\ell^b)$, hence
                              
\begin{corollary}
A 2-server PIR scheme with $t$-bit queries and $\ell$-bit answers
has communication
$
C = 2(t + \ell) = \Omega\left(n^{1/(b+1)}\right).
$
\end{corollary}
For $b=1$ this gives $C=\Omega(\sqrt{n})$, which is achieved by  
the square scheme of Section~\ref{ssecsquarescheme}.
For $b=3$ we get $C=\Omega(n^{1/4})$, which is close to the $C=O(n^{1/3})$ 
of the cube scheme.
As in Theorem~\ref{mtwoqueryparity}, we can get the better bound 
$
t = \Omega(n(\eps-\eta)^2/{\ell \choose b})
$
for PIR schemes where
the user just outputs the parity of $b$ bits from each answer.
All known non-trivial PIR schemes have this property.

\subsection{Weak lower bounds for general 2-server PIR}

The previous lower bounds on the query length of 2-server PIR schemes
were significant only for protocols that use few bits from each answer.
Here we slightly improve the best known bound of $4.4\log n$~\cite{kerenidis&wolf:qldcj}
on the overall communication complexity of 2-server PIR schemes,
by combining our Theorem~\ref{pirlarger} and Theorem~6 of Katz and
Trevisan~\cite{katz&trevisan:ldc}.
We restate their theorem for the PIR setting, assuming for simplicity that
$\eps=1/2$ and $\eta=0$.
                                                                                                                               
\begin{theorem}[Katz \&~Trevisan]\label{KTpir}
Every 2-server PIR with $t$-bit queries and $\ell$-bit answers has
$
t \geq 2\log(n/\ell) - O(1).
$
\end{theorem}
                                                                                     
We now prove the following lower bound on the
total communication $C=2(t+\ell)$ of any 2-server PIR scheme
with $t$-bit queries and $\ell$-bit answers:                                                                                                                  
\begin{theorem}
Every 2-server PIR scheme has 
$
C \geq \left(5-o(1)\right)\log n.
$
\end{theorem}                                                                                                                       
\begin{proof}
We distinguish three cases, depending on the answer length.
Let $\delta=\log\log n/\log n$.\\
{\bf case 1:} $\ell \leq (0.5-\delta)\log n$.
Theorem~\ref{pirlarger} implies $C\geq t=\Omega(n^{2\delta})=\Omega((\log n)^2)$.\\
{\bf case 2:} $ (0.5-\delta)\log n < \ell < 2.5\log n$.
Then from Theorem~\ref{KTpir} we have\\ 
$
C = 2(t+\ell) > 2\left(2\log(n/(2.5\log n)) - O(1) + \left(0.5-\delta\right)\log n\right)
= \left(5-o(1)\right)\log n.
$\\
{\bf case 3:} $\ell \geq 2.5\log n$.
Then $C=2(t+\ell)\geq 5\log n$.
\end{proof}

\section{Conclusion and Future Work}

Here we improved the best known lower bounds on the length
of 2-query locally decodable codes and the communication complexity 
of 2-server private information retrieval schemes.
Our bounds are significant whenever the decoder uses only few bits 
from the two query answers, even if the alphabet (LDC case) or answer 
length (PIR case) is large.
This contrasts with the earlier results of Kerenidis and 
de Wolf~\cite{kerenidis&wolf:qldcj}, which become trivial 
for logarithmic alphabet or answer length, and those of 
Goldreich et al.~\cite{gkst:lowerpir}, which only apply to \emph{linear} schemes.

Still, general lower bounds without constraints on alphabet or answer size
completely elude us. Clearly, this is one of the main open questions 
in this area. Barring that, we could at least improve the dependence
on $b$ of our current bounds.
For example, a PIR lower bound like $t=\Omega(n/\ell^{\ceil{b/2}})$ might 
be feasible using some additional quantum tricks.  Such a bound 
for instance implies that the total communication is $\Omega(n^{1/3})$ 
for $b=3$, which would show that the cube scheme of~\cite{cgks:pir}
is optimal among all schemes of probe complexity 3.
Another question is to obtain strong lower bounds for
the case of $k\geq 3$ queries or servers. For this case, no superpolynomial
lower bounds are known even if the alphabet or answer size is only one bit.

\end{document}